\documentclass[preprint,final,5p,times,twocolumn]{elsarticle}
\usepackage{amssymb}
\usepackage{graphicx}
\usepackage{multirow}
\usepackage{color}
\usepackage{amsthm}
\usepackage{amsmath}
\usepackage{widetext}
\usepackage{amssymb}
\usepackage{float}
\usepackage[utf8x]{inputenc}
\usepackage[numbers]{natbib}
\usepackage[colorlinks=true,citecolor=blue]{hyperref}
\journal{Physics Letters B}
\begin{document}
\begin{frontmatter}
\title{Constraining the relativistic mean-field models from PREX-2 data: Effective forces revisited}
\author{Jeet Amrit Pattnaik$^{1}$}
\author{R. N. Panda$^{1}$}
\author{M. Bhuyan$^{2,3}$}
\ead{bunuphy@um.edu.my}
\author{S. K. Patra$^{4,5}$}
\address[1]{Department of Physics, Siksha $'O'$ Anusandhan, Deemed to be University, Bhubaneswar-751030, India}
\address[2]{Center of theoretical and Computational Physics, Department of Physics, University of Malaya, Kuala Lumpur, 50603, Malaysia}
\address[3]{Institute of Research and Development, Duy Tan University, Da Nang 550000, Vietnam}
\address[4]{Institute of Physics, Sachivalya Marg, Bhubaneswar-751005, India}
\address[5]{Homi Bhabha National Institute, Training School Complex, Anushakti Nagar, Mumbai 400094, India}
\begin{abstract}
Based on the current measurement of the neutron distribution radius ($R_n$) of $^{208}$Pb through the PREX-2 data, we re-visited the recently developed G3 and IOPB-I force parameter by fine-tuning some of the specific couplings within the relativistic mean-field model. The $\omega-\rho-$mesons coupling $\Lambda_{\omega}$ and the $\rho-$meson coupling $g_{\rho}$ are refitted to reproduce the experimental neutron radius of $^{208}$Pb without compromising the bulk properties of finite nuclei and infinite nuclear matter observables. The modified parameter sets are applied to calculate the gross properties of finite nuclei for a few double closed-shell nuclei and further used to obtain the various infinite nuclear matter observables at saturation. In addition to these, the force parameters are adopted to calculate the properties of high isospin asymmetry dense system such as neutron star matter and tested for the validation for the constraint from GW170817 binary neutron star merger events. The tuned forces are predicting relatively good results for finite and infinite nuclear matter systems and the current limitation on neutron radius from PREX-2. A systematic analysis using these two refitted parameter sets over the nuclear chat will be communicated shortly.    
\end{abstract}
\end{frontmatter}

\section{Introduction}
\label{sec1}
\noindent
The physics of low-mass neutron star (NS),  the supernovae explosion, and the formation of new elements are governed by the same set of parameters, which predict the properties of finite nuclei, especially the neutron skin-thickness ($\Delta R_{np}$) \cite{adhikari21,reed21}. Since the proton is a charged particle, the precise measurement of its radius $R_p$ is possible. However, an accurate determination of the neutron distribution inside a finite nucleus suffers significant uncertainties \cite{thiel19,abrahamyan12,horo12}. The exact measurement of neutron radius $R_n$ and/or neutron skin-thickness $\triangle {R_{np}}=R_n-R_p$ will be helpful to calibrate many theoretical models in terms of the quantity directly related to the isospin asymmetry.  It is worth mentioning that the symmetry energy $J$ and its slope parameter $L$ are crucial entities to understand the equation of state (EoS) and vice-versa. A lot of attempts have been initiated to fix its value and its correlation with other physical quantities \cite{agrawal12,agrawal13,agrawal12a,centelles09,centelles11,centelles14}. For example, the strong correlations of neutron-skin thickness in $^{208}$Pb nucleus with various neutron star properties within the relativistic mean-field (RMF) models have described either in terms of the coupling of the isovector-vector $\rho$- meson with the nucleons \cite{lala97} and/or the cross-coupling of isoscalar-vector $\omega$- with the isovector-vector $\rho$- mesons \cite{estal01b,todd05,bhu13}. \\

The pressure of the neutron-rich matter normally pushing the surface tension in the atomic nucleus and plays the same against gravity in a neutron star. Even the difference in their magnitude differ in a large scale but share a common origin, which is sensitive to the equation of state (EoS). The precise measurement of neutron radius $R_n$ is supposed to be possible by using the parity violating weak neutral interaction at Thomas Jefferson National Accelerator Facility (JLab) termed as Lead Radius Experiment (PREX) \cite{donnelly89,horowitz01}. Based on this principle, the PREX-I result for $R_n$ of $^{208}$Pb is reported in Ref. \cite{abrahamyan12} with $R_n=5.78^{+0.16}_{−0.18}$ fm and a neutron skin-thickness of $\triangle{R_{np}}= R_n-R_p = 0.33^{ +0.16}_{-0.18}$ fm. To reduce the uncertainty further, the PREX-2 result is published with the neutron skin thickness as $\triangle{R_{np}=0.283\pm{0.071}}$ fm \cite{adhikari21} and point neutron distribution radius $R_n=5.727\pm{0.071}$ fm [knowing the precise value of point proton distribution radius as $R_p=5.444$ fm \cite{frois77,jones14} with the  corresponding charge radius $R_{ch}=5.0$ fm \cite{ong10}]. The precise measurement of $R_n$ (or $\Delta R_{np}$) allows refitting of the relevant parameters with the theoretical models to reproduce the properties of finite nuclei in parallel to describe the properties of neutron star. \\

Tempted by the PREX-2 result \cite{adhikari21}, which has $\sim $1 $\%$ accuracy, Reed {\it et al.} \cite{reed21} employed the property of strong correlation of neutron skin thickness $\triangle{R}_{np}$ with the slope parameter of symmetry energy $L$ and constrained it's value to $L = 106 \pm 37$ MeV \cite{centelles09,centelles11,brown00,furnstahl02}.  And the symmetry energy $J$ is fixed to be in the range $J=38.1\pm{4.7}$ MeV by using specific sets of relativistic mean-field (RMF) parametrizations. These values of $L$ and $J$ are surely larger than the presently settled values obtained either from theoretical models or from various experimental measurements \cite{zhang13,hebeler13,drischler20,hagen16,chen10,steiner12,gandolfi14,rocamaza15}. The precise measurement of the neutron skin-thickness of $^{208}$Pb by Adhikari {\it et al.} \cite{adhikari21} and the interesting results of Reed {\it et al.} \cite{reed21} motivate to revisit the recent RMF parameter sets, namely, G3 and IOPB-I by tunning the essential couplings that rarely affect the global properties of the infinite nuclear matter and finite nuclei. It is worth mentioning that, in general, these two forces reproduce well the known experimental data for finite nuclei and the properties of NS and the gravitational waves strain in binary neutron star mergers \cite{kumar17,kumar18,kumar17b}. The two couplings $\Lambda_{\omega}$ and $g_{\rho}$ of the RMF Lagrangian are taken as the tuning parameters to reproduce the recent experimental $R_n$ for $^{208}$Pb. The detailed procedure is highlighted in the subsequent sections. \\

The paper is arranged as follows: After a brief introduction in Section \ref{sec1}, a short description of the formalism is given for relativistic mean-field formalism in Section \ref{theory}. Since the RMF is already a standard theory, we only outlined the essential ingredients needed for the discussions. Section. \ref{parchose} is created for the justification of the parameter chosen. In Section \ref{results}, we have discussed our results and compared with the empirical/experimental data. Finally, a concluding remark is given in Section \ref{summary}. 
\section{Relativistic mean field (RMF) Model}
\label{theory}
The nonlinear relativistic mean-field lagrangian density is constructed by the interaction of nucleons with the well known $\sigma-$, $\omega-$, $\rho-$, $\delta-$ and photon fields, generated by $\sigma-$, $\omega-$,$\rho-$, $\delta-$mesons and protons, respectively.  The self and cross-couplings among the mesons are also included in the extended relativistic mean-field (E-RMF) theory, which is evolved in the framework of effective field theory motivated by naive dimensional analysis (NDA) and naturalness concept. Taking into this E-RMF, NDA, and naturalness criteria, the G3 and IOPB-I force parameters are designed, and the E-RMF Lagrangian density is written as, \cite{kumar17,kumar18,frun96,frun97,muller,serotijmpe}:
\begin{widetext}
\begin{eqnarray}
{\cal E}({r})&=&\sum_{\alpha=p,n} \varphi_\alpha^\dagger({r})\Bigg\{-i \mbox{\boldmath$\alpha$} \!\cdot\!\mbox{\boldmath$\nabla$}+\beta \bigg[M-\Phi (r)-\tau_3 D(r)\bigg]+ W({r})+\frac{1}{2}\tau_3 R({r})+\frac{1+\tau_3}{2} A({r})-\frac{i\beta\mbox{\boldmath$\alpha$}}{2M}\!\cdot\!\bigg(f_\omega\mbox{\boldmath$\nabla$}W({r})
\nonumber\\
&&
+\frac{1}{2}f_\rho\tau_3 \mbox{\boldmath$\nabla$}R({r})\bigg)\Bigg\} \varphi_\alpha(r)+\left(\frac{1}{2}
+\frac{\kappa_3}{3!}\frac{\Phi({r})}{M}+\frac{\kappa_4}{4!}\frac{\Phi^2({r})}{M^2}\right)
\frac{m_s^2}{g_s^2}\Phi^2({r})-\frac{\zeta_0}{4!}\frac{1}{g_\omega^2 }W^4({r})+\frac{1}{2g_s^2}\left(1+\alpha_1\frac{\Phi({r})}{M}\right) \bigg(
\mbox{\boldmath $\nabla$}\Phi({r})\bigg)^2
\nonumber\\
&&
-\frac{1}{2g_\omega^2}\left( 1 +\alpha_2\frac{\Phi({r})}{M}\right)\bigg(\mbox{\boldmath$\nabla$} W({r})\bigg)^2-\frac{1}{2}\left(1+\eta_1\frac{\Phi({r})}{M}+\frac{\eta_2}{2} \frac{\Phi^2({r})}{M^2}\right)\frac{m_\omega^2}{g_\omega^2} W^2({r})-\frac{1}{2e^2} \bigg( \mbox{\boldmath $\nabla$} A({r})\bigg)^2-\frac{1}{2g_\rho^2} \bigg( \mbox{\boldmath $\nabla$} R({r})\bigg)^2
\nonumber\\
&& 
-\frac{1}{2} \left( 1 + \eta_\rho \frac{\Phi({r})}{M}\right)\frac{m_\rho^2}{g_\rho^2} R^2({r}) -\Lambda_{\omega}\bigg(R^{2}(r)\times W^{2}(r)\bigg)+\frac{1}{2 g_{\delta}^{2}}\left( \mbox{\boldmath $\nabla$} D({r})\right)^2+\frac{1}{2}\frac{{m_{\delta}}^2}{g_{\delta}^{2}}D^{2}(r)\;.
\label{eds}
\end{eqnarray}
\end{widetext}
\noindent
Here $\Phi = g_s\sigma$, $W = g_\omega \omega$, $R$ = g$_\rho\vec{\rho}$ and $D=g_\delta\delta$ are the re-defined fields, with their coupling constants $g_\sigma$, $g_\omega$, $g_\rho$, $g_\delta$, and their  masses are $m_\sigma$, $m_\omega$, $m_\rho$ and $m_\delta$, respectively  for $\sigma$, $\omega$, $\rho$, and  $\delta$ mesons, and $\frac{e^2}{4\pi}$ is the photon coupling constant. From the effective-RMF energy density [Eq. (\ref{eds})], a set of coupled differential equations for finite nuclei (equation of motion) and the expression for pressure and energy (equation of state) for infinite nuclear matter are obtained using the Euler-Lagrange equation and the energy-momentum tensor, respectively \cite{kumar18}. The scalar and vector densities are
\begin{eqnarray}
\rho_s(r)&=&\sum_\alpha \varphi_\alpha^\dagger({r})\beta\varphi_\alpha, \label{scaden}  \\
\rho_v(r)&=&\sum_\alpha \varphi_\alpha^\dagger({r})\tau_{3}\varphi_\alpha\label{vecden},
\end{eqnarray}
respectively. A detail numerical evaluation is available in Refs. \cite{estal01b,kumar17,kumar18,estal01a}.

\section{Parameter chosen}\label{parchose}
The inception of the relativistic mean-field model goes back to 1955  when it is proposed a classical field theory and the relativistic formulation of Nuclear Force \cite{teller55,duerr56}. Later on, a proper mathematical formulation is given to the model by Miller and Green \cite{green72}, Brockmann \cite{brockmann78} assuming a scalar and vector interaction potential. Finally, Walecka \cite{walecka74}, Serot and collaborators \cite{serot81,walecka86} extended the formalism to the various domain of finite and infinite nuclear systems. Each extension in the model shows that every interaction corresponds to a particular property of the nuclear potential. For example, the $\sigma-$meson mainly responsible for the strong attraction at the intermediate range of the nuclear force, but the self-interaction of this scalar meson produces a weak repulsion at long-range \cite{boguta77,biswal15}. Furthermore, the self-interaction of the $\sigma-$meson allow to constrain the incompressibility of nuclear matter at nuclear saturation to the empirical range, $K_{\infty} = 210 \pm 20$ MeV \cite{blaizot80}. The $\omega-$meson responsible for the strong hard-core repulsion of the nuclear potential and its self-interaction generates attraction at a concise range and makes the nuclear equation of state (EoS) softer \cite{bodmer91,gmuca92,sugahara94}. A detailed discussion can be found on the influence of various interactions in Refs. \cite{estal01a,biswal15}.\\

At present, it is clear that each interaction in the Lagrangian density leads to physical property, either the finite nuclear or infinite nuclear matter system. Thus, the effective field theory with the NDA and naturalness approach added to E-RMF formalism, and all possible important interactions are included in the framework. In this way, G1 and G2 parameter sets are proposed in Refs. \cite{frun96,frun97}. The effects of $\delta-$meson to the nuclear potential is realised in \cite{singh14a,singh14b}. The $\delta-$meson interaction along with the couplings of G1 and G2 is the product of G3 set  \cite{kumar17}. A shorter version of this parameter known as IOPB-I is also designed \cite{kumar18}.  In the present analysis, the G3 and IOPB-I parameter sets are revisited to constrain the PREX-2 data for neutron skin-thickness $\Delta R_{np}$ and vice-versa. In other words, the G3 and IOPB-I parameter sets are readjusted by considering the PREX-2 data and applied these new/modified parameterizations for the structural analysis of finite nuclei and the properties of infinite nuclear matter, including the neutron star.\\

\begin{table}
\caption{The old/original G3(O) \cite{kumar17}, IOPB-I(O) \cite{kumar18} and their new/modified sets G3(M), IOPB-I(M) are listed. The mass of nucleon $M$ is 939 MeV. The dimension of $k_3$ is fm$^{-1}$, and all other coupling constants are dimensionless. The exact values of $\Lambda_{\omega}=$0.02112981 and $g_{\rho}=$10.961218044 for G3(M) and $\Lambda_{\omega}=$0.01475398 and $g_{\rho}=$10.0907956536 for IOPB-I(M). The (O) and (M) stand for old/original and new/modified/revisited versions of the parametrizations and their corresponding calculations.}
\begin{tabular}{ccccc}
\hline
\hline
\multicolumn{1}{c}{Parameter}
&\multicolumn{1}{c}{G3(O)}
&\multicolumn{1}{c}{G3(M)}
&\multicolumn{1}{c}{IOPB-I(O)}
&\multicolumn{1}{c}{IOPB-I(M)}\\
\hline
$m_{s}/M$  &  0.559&  0.559&0.533&0.533  \\
$m_{\omega}/M$&  0.832&  0.832&0.833& 0.833 \\
$m_{\rho}/M$&  0.820&  0.820&0.812& 0.812 \\
$m_{\delta}/M$ &   1.043&   1.043&0.0& 0.0 \\
$g_{s}/4 \pi$  &   0.782&  0.782 &0.827&0.827 \\
$g_{\omega}/4\pi$ & 0.923&  0.923&1.062&1.062 \\
$g_{\rho}/4 \pi$& 0.962 &0.872&0.885&0.803 \\
$g_{\delta}/4 \pi$  & 0.160&0.160& 0.0&0.0 \\
$k_{3} $   & 2.606 &    2.606&1.496&1.496 \\
$k_{4}$  & 1.694 & 1.694&-2.932& -2.932 \\
$\zeta_{0}$  &  1.010&  1.010  &3.103& 3.103\\
$\eta_{1}$  & 0.424&  0.424 &0.0&0.0 \\
$\eta_{2}$  &  0.114&  0.114  &0.0&0.0 \\
$\eta_{\rho}$  & 0.645&  0.645& 0.0 &0.0 \\
$\Lambda_{\omega}$& 0.038&0.021&0.024&0.015\\
$\alpha_{1}$  &   2.000&   2.000&0.0&0.0 \\
$\alpha_{2}$  &  -1.468&  -1.468&0.0 &0.0 \\
$f_\omega/4$  &  0.220&  0.220&0.0&0.0 \\
$f_\rho/4$  &   1.239& 1.239&0.0& 0.0\\
$\beta_\sigma$  & -0.087& -0.087& 0.0& 0.0 \\
$\beta_\omega$& -0.484& -0.484& 0.0&0.0  \\
\hline
\hline
\end{tabular}
\label{table1}
\end{table}
\begin{table*}
\caption{The binding energy (BE in MeV) and neutron distribution radius ($R_n$ in fm) for some selected spherical nuclei with modified G3(M), IOPB-I(M) parameter sets along with the old/original G3(O), and IOPB-I(O) predictions [in parentheses]. The charge radius $R_{ch}$ is obtained by adopting the finite size effect of the nucleon, i.e., $R_{ch} = \sqrt{R_{p}^2 + 0.64}$ fm. The experimental data are taken from Refs. \cite{adhikari21,seif,kra99,trz01,angeli} and reference therein. Follow the texts for more details.  \\}
\renewcommand{\tabcolsep}{0.06cm}
\renewcommand{\arraystretch}{1.3}
\begin{tabular}{ccccccccccc}
\hline\hline 
\multicolumn{1}{c}{} & \multicolumn{3}{c}{Binding Energy} & \multicolumn{3}{c}{$R_{ch}$} & \multicolumn{3}{c}{$R_{n}$} \\
Nucl. & G3(M) [G3(O)]& IOPB-I(M) & Expt. & G3(M) [G3(O)]& IOPB-I(M) & Expt. & G3(M) [G3(O)]& IOPB-I(M) & Expt. \\
&&[IOPB-I(O)]&&&[IOPB-I(O)]&&&[IOPB-I(O)] & \\
\hline
$^{16}$O  & 128.59[128.59] & 127.63[127.63] & 127.62 &2.706[2.706] & 2.704[2.705]& 2.699& 2.613[2.612] &2.575[2.575] & ---   \\
$^{40}$Ca&342.46[342.46]&343.07[343.07]&342.05& 3.458[3.458]& 3.457[3.458]& 3.477& 3.351[3.351]& 3.322[3.321]&$3.306_{-1.0}^{+0.05}$\\
$^{48}$Ca&416.99[416.18] & 415.04[414.62]&415.96& 3.461[3.466]& 3.436[3.441]& 3.477& 3.621[3.613]& 3.593[3.583]&$3.499_{-0.05}^{+0.05}$\\
$^{90}$Zr&783.25[782.91]& 782.25[782.23]&783.81& 4.271[4.275]& 4.249[4.253]& 4.269& 4.319[4.312] &4.297[4.289]&$4.283_{-0.02}^{+0.02}$ \\
$^{116}$Sn&985.67[985.67]& 986.65[986.85]&988.66& 4.630[4.634]& 4.615[4.621]& 4.625& 4.715[4.704] & 4.701[4.688]& $4.692_{-0.05}^{+0.05}$ \\
$^{132}$Sn&1105.43[1103.54]& 1103.06[1102.42]& 1102.72& 4.725[4.731]& 4.699[4.705]& 4.709& 4.970[4.948] & 4.965[4.941] &$4.880_{-0.04}^{+0.04}$\\
$^{208}$Pb& 1636.43[1635.51] & 1636.42[1636.87] & 1636.43 &5.336[5.541]& 5.516[5.521]& 5.501& 5.717[5.694] &5.721[5.696] &$5.727_{-0.071}^{+0.071}$ \\
$^{304}$120&2131.41[2131.02] &2134.37[2135.71]& 
---& 6.336[6.338]& 6.327[6.329]&---& 6.487[6.460] &6.507[6.478]&---\\
\hline \hline
\end{tabular}
\label{tab2}
\end{table*}

\section{Results and discussions}
\label{results} 
The cross-coupling of $\rho-$meson with $\sigma-$ and $\omega-$mesons allows for varying neutron skin thickness in heavy mass nucleus like $^{208}$Pb \cite{pika05}. This coupling $\Lambda_{\omega}$ also matters a lot for the giant resonances, like monopole and quadrupole \cite{patra10}. The $\rho-$meson coupling takes care of the neutron-proton asymmetry in the system. Therefore, the apparent choice of the minimal tuning of parameters is the $\Lambda_{\omega}$ and $g_{\rho}$. In addition to this nature of $\Lambda_{\omega}$ and $g_{\rho}$, the other suitable behavior for the modifications are that the binding energy (BE) of the asymmetric nuclear system increases with $\Lambda_{\omega}$. In contrast, the binding energy decreases with the increasing value of $g_{\rho}$. On the other hand, the neutron distribution radius $R_n$ decreases with the increase of $\Lambda_{\omega}$, while the $R_n$ remains almost unchanged with $g_{\rho}$. In Table \ref{table1}, the old/original G3(O), IOPB-I(O) and their new/modified  values G3(M), IOPB-I(M) are listed. It is to be noted that only the values of $\Lambda_{\omega}$ and $g_{\rho}$ are modified and analysed the effects on various properties of finite and infinite nuclear systems. \\

\begin{figure}[H]
\includegraphics[width=1.0 \columnwidth]{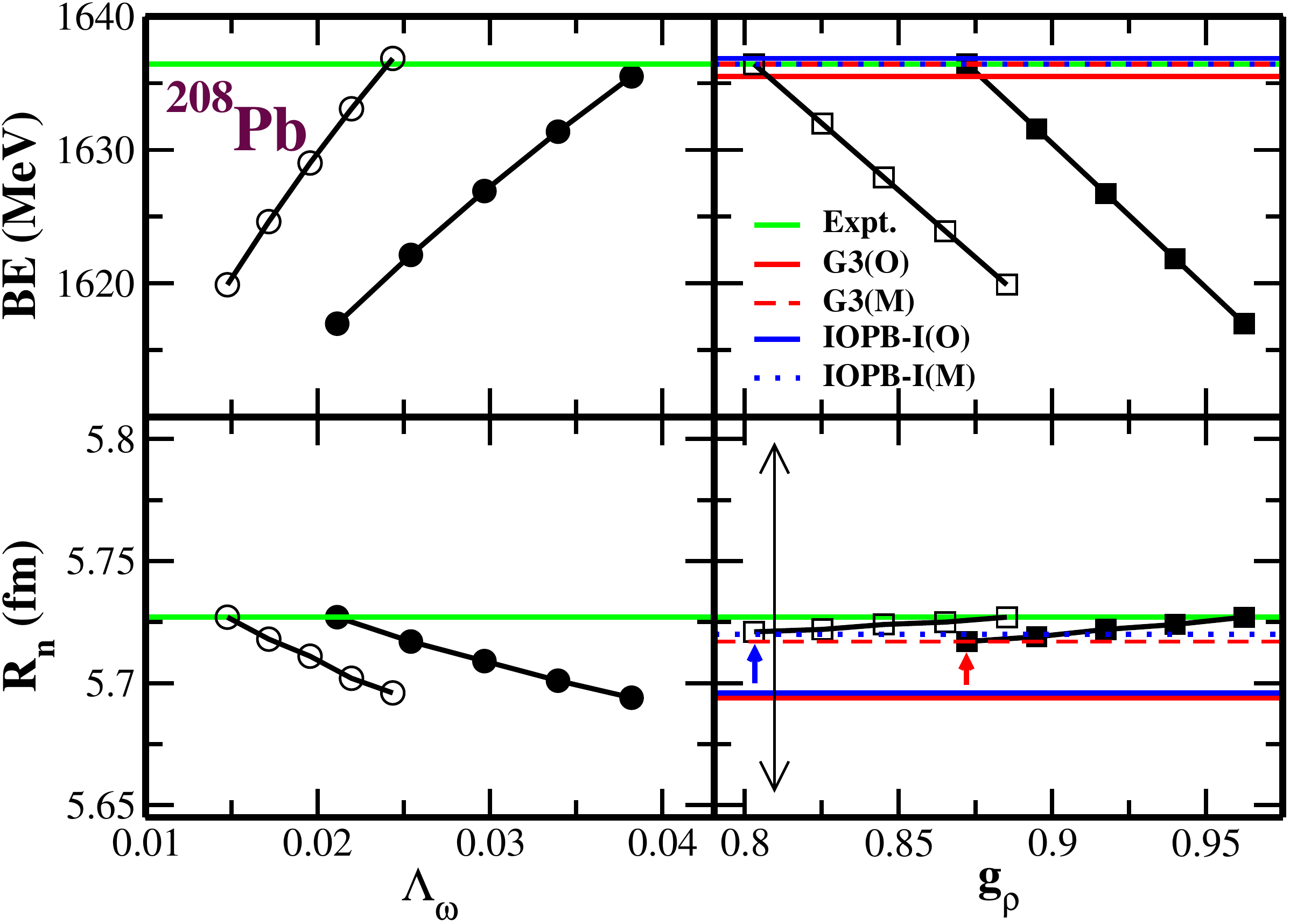}
\caption{(Color online) The binding energy (BE in MeV) and neutron distribution radius $(R_n$ in fm) with old/original G3(O) (red) and new/modified G3(M) (dotted red) forces for $^{208}$Pb as a function of $\omega-\rho-$fields coupling constant $\Lambda_{\omega}$ and $\rho-$meson coupling $g_{\rho}$. (For IOPB-I(O) and IOPB-I(M) it is blue and dotted blue respectively). The upper panel is the binding energy, and the lower panel is the neutron distribution radius. We have also shown the changes of BE and $R_n$ corresponding to both G3(M) (black filled) and IOPB-I(M) (black empty). The experimental data (green) \cite{adhikari21,wang2017} are also given. The double arrow represents for error bar and the single arrow for the better visibility of $R_n$.}
\label{fig1}
\end{figure}

The effects of $\Lambda_{\omega}$ and $g_{\rho}$ on binding energy and $R_n$ with $\Lambda_{\omega}$ and $g_{\rho}$ are shown in Fig. \ref{fig1}. We calculate the BE and $R_n$ with variation of $\Lambda_{\omega}$ from 0.03821181 to 0.02112981 without changing $g_{\rho}$ for G3 parameter set. As stated above, we find the increase of binding energy with $\Lambda_{\omega}$. Then, we change the values of $g_{\rho}$ in the range 12.0947046734 $-$10.961218044. In this case, the binding energy decreases with the increase of $g_{\rho}$ without affecting the $R_n$. The experimental binding energy \cite{wang2017} has been shown in the green line. By fixing the experimental data of BE and $R_n$, we calibrated the $\Lambda_{\omega}$ and $g_{\rho}$ combinations, which is termed as new/modified G3(M) parameter set, and the ($\Lambda_{\omega},g_{\rho}$) as (0.02112981 \& 10.961218044), which are tabulated in Table \ref{table1}. We follow the same procedure for the IOPB-I parameter set and found the new/modified values of $\Lambda_{\omega}$ and $g_{\rho}$ as $\Lambda_{\omega}=$0.01475398 and $g_{\rho}=$10.0907956536. The results for binding energy and neutron distribution radius are also shown in  Fig. \ref{fig1}. \\

\begin{table}
\caption{The nuclear matter (NM) properties (in MeV) at saturation density ($\rho_0$ in fm$^{-3}$) and the neutron star (NS) properties, such as, maximum mass (M in solar mass unit) and  radius (R km), are obtained by using the old/original G3(O), IOPB-I(O) and new/modified G3(M), IOPB-I(M) parameter sets. The tidal Love number $k_2$, tidal deformability ($\lambda$) and dimensionless tidal deformability $(\Lambda$) at 1.4 solar mass are also presented. The experimental values for both NM and NS quantities are given for comparison, wherever available.  \\}
\renewcommand{\tabcolsep}{0.07cm}
\renewcommand{\arraystretch}{1.1}
\begin{tabular}{cccccc}
\hline\hline
NM & G3 & G3 & IOPB-I & IOPB-I & Expt.\\
NS & (O) & (M) & (O) & (M) & \\
\hline
$J$      &  31.842  &       33.063  &   33.355   &    34.831&69 - 143 \cite{reed21}\\
$L$      &  49.317  &       65.530  &   63.700   &    78.947& 33.4 - 42.8 \cite{reed21}\\
$K_{sym}$  & -106.07 &     -113.99  &  -36.60   &   -61.46 & - (174 - 31) \cite{Zimmerman_2020}\\  $Q_{sym}$  &  915.47 &      525.46  &  859.90   &   559.57 & ---\\
$K_{\infty}$      &  243.97 & 243.97  &  222.33 & 222.33 & 220 - 260 \cite{Colo_2014}\\
$K_{asy}$  & -401.98 & -507.17  & -418.80 &  -535.14 & ---\\        
$Q_0$    & -466.61 & -466.61 &  -96.67    &  -96.67 & ---\\
$K_{tau}$  & -307.65 & -381.84  & -391.10 & -500.81 & - (840 - 350) \\
&&&&& \cite{Stone_2014,Pearson_2010,TLi_2010}\\
$\rho_{0}$    &  0.148   & 0.148   &  0.149  &  0.149&0.148 - 0.185 \cite{Bethe_1971}  \\         
$K_{sat2}$ &  -307.65& -381.84  & -391.10    & -500.81& ---\\
$M_0$    &  2460.98 & 2460.98 & 2571.31 & 2571.31 & ---\\
E/A  &  -16.024 &      -16.024  &  -16.105    &  -16.105& - (15.0 - 17.0) \cite{Bethe_1971}\\
&&&&&\\
\hline\hline
M     &  1.99 &    1.99  &    2.14 &2.14&$1.97_{-0.04}^{+0.04}$\cite{demorest},\\
 & & & & &  $2.14_{-0.09}^{+0.10}$\cite{cromatie},\\
  & & & & &  $2.01_{-0.04}^{+0.04}$ \cite{antoniadis}\\
R   &   10.81 &    10.87 &    11.80   & 11.85&$13.02_{-1.06}^{+1.24}$, $12.71_{-1.19}^{+1.14}$\\
&&&&& \cite{NICER}\\
$\Lambda$&  464.63  &  498.23  &  689.62 &   727.43&70-580 \cite{abbott,ligo17,ligo18}	\\	
$\lambda_2$    & 2.61 & 2.81&  3.87 &  4.11&---\\
$k_2$  &   0.09 &  0.09 &  0.11  &  0.11&---\\
\hline\hline 
\end{tabular}
\label{tab3}
\end{table}

Once constraining the values of $\Lambda_{\omega}$ and $g_{\rho}$ to the experimental data of $R_n$ and the binding energy for $^{208}$Pb, we calculate the BE and $R_n$ for some of the spherical nuclei. The calculated results are displayed in Table \ref{tab2} along with the experimental data  \cite{adhikari21,wang2017,seif,kra99,trz01,angeli} . As expected, the BE and $R_n$ remain unchanged for symmetric nuclei where the number of neutrons is equal to the number of protons, i.e., N=Z nuclei ($^{16}$O and $^{40}$Ca). The neutron distribution radius changes considerably for all other nuclei, with a marginal influence on the binding energy. The nuclear matter quantities such as binding energy per particle for symmetric nuclear matter (E/A), the isospin dependence observable symmetric energy $J$, slope parameter $L$, surface symmetric energy coefficient $K_{sym}$, skewness parameter $Q_{sym}$, incompressibility $K_{\infty}$ are calculated using new/modified G3(M) and IOPB-I(M) parameter sets. Furthermore, the incompressibility of asymmetric nuclear matter coefficient $K_{asy}$, $Q_0 = 27\rho^3\frac{\partial^3 \cal{E}}{\partial {\rho}^3}$ in symmetric nuclear matter,  isospin asymmetric coefficient  $K_{tau}$, central density $\rho_0$, incompressibility of second-order at saturation density $K_{sat2}$ and slope of the incompressibility $M_0$ are also estimated new/modified G3 and IOPB-I parameter sets. The calculated nuclear matter (NM) properties along with Neutron star (NS) properties from the new/modified G3(M) and IOPB-I(M) are listed in Table \ref{tab3} with the old/original G3(O), and IOPB-I(O) estimates and the experimental/empirical values. \\

Constraining the $\Lambda_{\omega}$ and $g_{\rho}$ parameters with respect to the PREX-2 result of $R_n$, roughly, we find the $J$ and $L$ values as per within the limit of Reed {\it et al.} \cite{reed21}, i.e., $J=33.063-34.831$ MeV and $L=65.530-78.947$ MeV. Comparing the $J$ and $L$  of G3(M) and IOPB-I(M) sets, one can see that the IOPB-I(M) set is better suited with the prediction of Ref. \cite{reed21}. Also, G3(M) and IOPB-I(M) results are very much improved than their old/original values. Upon the analysis of $J$, $L$ and $K_{\infty}$, we find a significant enhancement in the symmetric energy coefficient $J$ and the slope parameter $L$ unaffecting the nuclear matter incompressibility. This is because the chosen parameters $\Lambda_{\omega}$ and $g_{\omega}$ do not strike the symmetric nuclear matter EoS. And, the symmetric energy coefficient is defined as $J(\rho)=\frac{1}{2}\left[\frac{\partial^2 {e}(\rho, \alpha)} {\partial \alpha^2}\right]_{\alpha=0}$, (where $e(\rho,\alpha)$ is the energy density obtained from the EoS and $\alpha\left(=\frac{\rho_n-\rho_p}{\rho_n+\rho_p}\right)$, with $\rho_n$ and $\rho_p$ are the neutrons and protons densities distribution). Practically, the value of $J$ at saturation density is obtained from the energy difference between the pure neutron matter (PNM) and the symmetric nuclear matter (SNM). Although the EoS of SNM does not depend on the $\rho-$meson coupling, but highly relies on asymmetric EoS.\\

\begin{figure}[H]
\centering
\includegraphics[width=1.0\columnwidth]{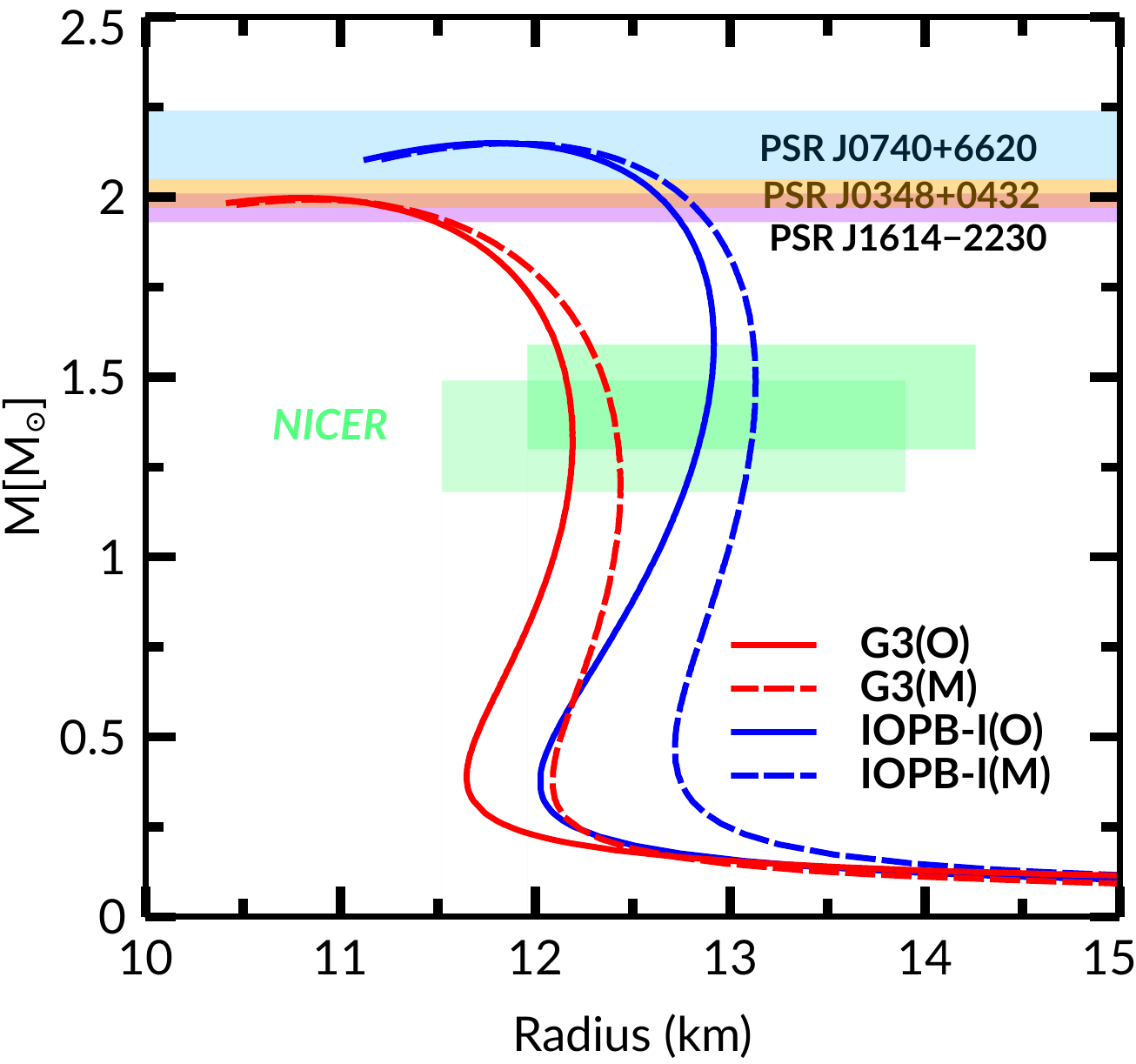}
\vspace{0.1 cm}
\caption{(color online) The mass radius (M-R) profile for the old/original G3(O), IOPB-I(O) and new/modified G3(M), IOPB-I(M) parameter sets. Some of the recent experimental data \cite{demorest,cromatie,antoniadis,NICER} are shown for comparison.}
\label{fig2}
\end{figure}
Finally, the new/modified forces G3(M) and IOPB-I(M) were applied to neutron star matter to determine mass $M$ and radius $R$. The $M-R$ profile with the old/original and the new/modified forces are shown in Fig. \ref{fig2} and also in Table \ref{tab3} as earlier discussed. The experimental observations for mass and possible radius \cite{demorest,cromatie} are also shown in the figure. It is interesting to note that the overall results for both new and old parameter sets are unchanged. The calculated mass and radius are obtained for both the old/original and new/modified parameter sets compiled in Table \ref{tab3}. The $M$ and $R$ with all the parameter sets are well within the recent measurements \cite{demorest,cromatie,antoniadis,NICER}. Moreover, we have calculated the highly discussed binary neutron star merger quantities \cite{ligo17,ligo18,bhu20} such as the Love number $k_2$, quadrupole tidal deformability $\lambda_2$ and the dimensionless tidal deformability $\Lambda_2$ for the new/modified G3(M) and IOPB-I(M) parameter set. In the above analysis, we find the new/modified version of G3 and IOPB-I are able to reproduce all the nuclear matter quanties and neutron star properties, including neutron star merger, compete with the old/original version along with the current PREX-2 data for neutron skin-thickness. More systematic analysis over the various region of the nuclear chart with systatic study of nuclear and star matter quantities will be communicated soon.  \\

\section{Summary and Conclusions}\label{summary}
In summary, we revisited the relativistic mean-field G3 and IOPB-I parameter sets, keeping the currently reported constraint on neutron radius of $^{208}$ by PREX-2. It is worth demonstrating that the precise measurement of neutron distribution radius discriminates various theoretical predictions. Hence it provides an opportunity to modify the force parameter or readjust the model with the implication of new interaction in the model. In this context, we have performed a minimal modification to the relevant couplings of recently developed G3 and IOPB-I parameter sets with E-RMF without compromising the predictions for finite nuclei and infinite nuclear matter. We did a fine-tuning of the $\omega-\rho$ cross-coupling $\Lambda_{\omega}$ along with the coupling of isovector-vector-meson ($g_{\rho}$) to reproduce the recent experimental $R_n$. The updated values of $\Lambda_{\omega}$ and $g_{\rho}$ for G3 set are (0.02112981 \& 0.872266017) and that of  IOPB-I set are (0.01475398 \& 0.803000004). We found the neutron distribution radius are 5.717 and 5.721 (in $fm$) meet the present experimental demand (R$_{n}^{Expt.}=5.727\pm{0.071}$) for modified G3(M) and IOPB-I(M), respectively. \\

The modified forces are used to calculate the bulk properties such as binding energy, root-mean-square charge distribution radius $R{ch}$ for $N\neq{Z}$ nuclei. Subsequently, the density and isospin-dependent nuclear matter parameters, such as symmetry energy $J$, slope $L$, and other specific observable, are also estimated, significantly favoring the experimental or other theoretical predictions. For example, the value of symmetry energy is 31.842 MeV for old/original G3(O), and after the modification, new G3(M) produces the value $J$ as 33.063 MeV. Similarly, the $J$ values are 33.355 and 34.831 MeV for IOPB-I(O) and IOPB-I(M), respectively. For both the parameter sets, the new values are closer to the limit set up by Reed {\it et al.} \cite{reed21}. Further, the forces are applied to calculate the properties of high isospin asymmetry dense systems such as neutron star matter and tested to validate the constraint from GW170817 binary neutron star merger events. We find relatively better predictions from the modified version of G3 and IOPB-I than the old version for finite and infinite nuclear matter systems. Hence, the re-calibration of the parameters in parallel generation of neutron radius of $^{208}$Pb constraint from PREX-2.
\\
\noindent 
{\bf  Acknowledgement:} One of the authors (JAP) is thankful to H. C. Das and Ankit Kumar for fruitful discussions \& suggestions and the Institute of Physics, Bhubaneswar, for providing computer facilities during the work. The SERB, Department of Science and Technology, Govt. of India, Project No. CRG/2019/002691 partly reinforces this work. MB acknowledges the support from FOSTECT Project No. FOSTECT.2019B.04, FAPESP Project No. 2017/05660-0, and the CNPq - Brasil. \\
\\

\end{document}